\documentclass[12pt]{article}
\usepackage{a4wide}
\usepackage{amsfonts}
\usepackage{fleqn,latexsym,amssymb,amsmath}
\usepackage{cite}
\usepackage[dvips]{graphicx}

\begin{document}
\renewcommand{\theequation}{\thesection.\arabic{equation}}
\numberwithin{equation}{section}


\begin{titlepage}
  \begin{flushright}
  {\small CAS-KITPC/ITP-022}
  \end{flushright}

  \begin{center}

    \vspace{20mm}

    {\LARGE \bf A Note on Exact Solutions and \\Attractor Mechanism \\
    \vspace{3mm}
    for Non-BPS Black Holes}

    \vspace{10mm}

    Rong-Gen Cai$^{\S\dag}$, Da-Wei Pang$^{\dag\ddag}$

    \vspace{5mm}
    {\small \sl $\S$ Kavli Institute for Theoretical Physics China(KITPC)\\
      at the Chinese Academy of Sciences} \\
    {\small \sl $\dag$ Institute of Theoretical Physics,\\
    Chinese Academy of Sciences} \\
    {\small \sl P.O.Box 2735, Beijing 100080, China} \\
    {\small \sl $\ddag$ Graduate University of the Chinese Academy of Sciences}\\
    {\small \sl YuQuan Road 19A, Beijing 100049, China\\}
    {\small \tt cairg@itp.ac.cn, pangdw@itp.ac.cn}
    \vspace{10mm}

  \end{center}

  \vskip 0.3 cm \centerline{\bf Abstract} \vspace{5mm}
  \noindent
  We obtain two extremal, spherically symmetric, non-BPS black
  hole solutions to 4D supergravity, one of which carries D2-D6
  charges and the other carries D0-D2-D4 charges. For the D2-D6
  case, rather than solving the equations of motion directly, we
  assume the form of the solution and then find that the
  assumption satisfies the equations of motion and the constraint.
  Our D2-D6 solution is manifestly dual to the solution presented
  in 0710.4967. The D0-D2-D4 solution is obtained by performing
  certain $[SL(2,{\bf Z})]^{3}$ duality transformations on the
  D0-D4 solution in 0710.4967.

\end{titlepage}
\setcounter{tocdepth}{2}
\tableofcontents
\section{Introduction}
Black holes have provided a variety of interesting research subjects
in recent years, one of which is the so-called ``attractor
mechanism''. It means that, in certain black hole background, the
moduli fields vary radially and ``get attracted'' to fixed values at
the black hole horizon, which depend only on the quantized charges
carried by the black hole. As a result, the entropy of the black
hole is given only in terms of the charges and is independent of the
asymptotic values of the moduli.

The attractor mechanism was firstly discovered in the mid 1990s in
the context of N=2 extremal black holes\cite{classic} and was
generalized to higher derivative theories in~\cite{cardoso}. The
supersymmetric attractors were the main focus at first, but later it
was realized that the attractor mechanism does not rely on
supersymmetry in~\cite{fgk}. Non-supersymmetric attractors have been
investigated extensively in recent years, see~\cite{non, kal1, kal2,
dst, other}. For reviews, see~\cite{review}.

It has been found that the non-BPS attractors share many interesting
properties with their BPS cousins, on condition that the non-BPS
black holes are extremal. Although we can study the non-BPS
attractors via the attractor equations, it is more useful to obtain
the solution for the moduli fields in the whole space. However, it
is rather complicated to obtain the exact solutions for non-BPS
black holes because the equations of motion are second-order
differential equations, rather than the first-order equations
appearing in BPS cases. Usually people deal with this difficulty by
making use of perturbative methods and numerical analysis, such as
in~\cite{non} and related references.

For STU black holes the situations have been better improved. It has
been shown in~\cite{WAS, WAS2, BLS, BD} that for the BPS case, we
can find solutions to the equations of motion for the moduli,
allowing us to obtain their values everywhere, by replacing the
charges appeared in the attractive values of the moduli with the
corresponding harmonic functions. A similar procedure has been
carried out in~\cite{kal2} for non-BPS black holes carrying D2-D6
charges. However, the solutions were still limited in the sense that
the moduli fields were taken to be purely imaginary.

Recently some interesting papers appeared~\cite{HK}~\cite{GLS},
which directly solved the equations of motion for the moduli in the
STU model carrying D0-D4 charges. The moduli and charges were set to
be equal in~\cite{HK} while more general cases were discussed
in~\cite{GLS}. Compared to the exact solutions in~\cite{kal2}, their
solutions for the moduli were complex.

We generalize the exact solutions in~\cite{HK} and~\cite{GLS} to STU
model carrying D2-D6 charges. It is difficult to solve the equations
of motion directly because unlike the D0-D4 case, the superpotential
contains a cubic term of the moduli fields. Instead of dealing with
the first order flow equations, we try to find the solution to the
second order equation of motion. By observing the exact solution
in~\cite{kal2}, which was manifestly dual to the existed D0-D4
system, we can assume that the D2-D6 solution is manifestly dual to
the solution in~\cite{GLS} and check if it satisfies the equations
of motion and the constraint. Fortunately, after a tedious
calculation we find that the assumption is correct.

It has been known that the symplectic invariance of special geometry
ensures that the Lagrangian has an $Sp(8, \bf Z)$ symmetry, which
reduces to $[SL(2,{\bf Z})]^{3}$ at the level of the the equations
of motion. Due to the $SL(2,\bf Z)$ duality, we can obtain new
solutions by making use of the ``seed solution''. We also obtain
D0-D2-D4 solution from the D0-D4 solution obtained in~\cite{GLS} by
some simple $SL(2,\bf Z)$ duality transformations.

The rest of the note is organized as follows: In Section 2 we give a
brief review of the attractor mechanism in STU model as well as the
main procedures and results in~\cite{GLS}. Then we assume the form
of the solution for STU model carrying D2-D6 charges with general
complex moduli and find that the solution satisfies both the
equations of motion and the constraint. Next we obtain D0-D2-D4
solution by dualizing the D0-D4 solution in~\cite{GLS} under
$[SL(2,{\bf Z})]^{3}$ symmetry. We summarize the results and discuss
some related topics in the final section.

\section{Preliminaries}
\subsection{The basic backgrounds}
Type IIA string theory compactified on a $CY_{3}$ manifold gives
$\mathcal{N}=2$ supersymmetry. The moduli fields belong to the
vector multiplets and hypermultiplets of the resulting low-energy
effective theory. The moduli in the vector multiplets are fixed
according to the attractor mechanism while the moduli in the
hypermultiplets play no role in the attractor mechanism. The
low-energy dynamics for the vector multiplets is completely
determined by a prepotential. If the Calabi-Yau manifold has
$h(1,1)=N$, we have $N$ vector multiplets and $N+1$ gauge fields,
where the additional gauge field is the graviphoton coming from the
gravity multiplet.  The leading order prepotential, ignoring any
$\alpha^{\prime}$ corrections, is given as
\begin{equation}
F=D_{ABC}\frac{X^{A}X^{B}X^{C}}{X^{0}},
\end{equation}
where $A,B,C=1,\cdots,N$. The intersection number $D_{abc}$ are
defined as
\begin{equation}
6D_{ABC}=\int_{CY_{3}}\alpha_{A}\wedge\alpha_{B}\wedge\alpha_{C},
\end{equation}
where $\alpha_{A}$ denote the integer basis for $H^{2}(CY_{3}, {\bf
Z})$.

Type IIA string theory admit D0, D2, D4 and D6 branes. D0 and D6
branes are electrically and magnetically charged with respect to the
graviphoton, while D2 and D4 branes are electrically and
magnetically charged with respect to the other $N$ gauge fields. One
can express the charge configuration collectively as
$(q_{0},q_{A},p^{0},p^{A})$, with $A=1,2,\cdots,N$. To be more
precise, let $\Sigma^{A}$ be a basis of 4-cycles dual to
$\alpha_{A}$ introduced above, and let $\hat{\Sigma}_{A}$ be a basis
dual to $\Sigma^{A}$. Then the magnetic charge carried by the D4
brane wrapping $\Sigma^{A}$ is $p^{A}$, while the electric charge
carried by the brane wrapping $\hat{\Sigma}_{A}$ is $q_{A}$.

The K\"{a}hler potential is given by
\begin{equation}
K=-\ln[i\sum\limits^{N}_{\Lambda=0}(\bar{X}^{\Lambda}F_{\Lambda}
-X^{\Lambda}\bar{F}_{\Lambda})],
\end{equation}
where $F_{\Lambda}\equiv\partial_{\Lambda}F$. If we choose the
inhomogeneous coordinates $z^{\lambda}=\frac{X^{\Lambda}}{X^{0}}$
and set the gauge $X^{0}=1$, the K\"{a}hler potential becomes
\begin{equation}
K=-\ln[-iD_{ABC}(z^{A}-\bar{z}^{A})(z^{B}-\bar{z}^{B})(z^{C}-\bar{z}^{C})].
\end{equation}
The superpotential is given by
\begin{equation}
W=q_{\Lambda}X^{\Lambda}-p^{\Lambda}F_{\Lambda}.
\end{equation}
Here in the $X^{0}=1$ gauge the superpotential becomes
\begin{equation}
\label{2eq6}
W=q_{0}+q_{A}z^{A}-3D_{ABC}z^{A}z^{B}p^{C}+p^{0}D_{ABC}z^{A}z^{B}z^{C}.
\end{equation}

\subsection{The STU model}
Now let us concentrate on the so-called STU model, which can be
interpreted in terms of type IIA string theory on a $T^{6}$ of the
form $T^{2}\times T^{2}\times T^{2}$. This model contains three
vectormultiplets, i.e. $N=3$. The prepotential and the K\"{a}hler
potential can be obtained immediately.
\begin{equation}
F=\frac{X^{1}X^{2}X^{3}}{X^{0}}.
\end{equation}
\begin{equation}
\label{2eq8}
K=-\ln[-i(z^{1}-\bar{z}^{1})(z^{2}-\bar{z}^{2})(z^{3}-\bar{z}^{3})]=
-\ln(8y^{1}y^{2}y^{3}),
\end{equation}
where we have rewritten $z^{a}=x^{a}-iy^{a}(a=1,2,3)$ for later
convenience. The metric and connection on the moduli space are given
by
\begin{equation}
\label{2eq9}
G_{a\bar{b}}=-\frac{\delta_{ab}}{(z^{a}-\bar{z}^{a})^2}=\frac{\delta_{ab}}{(2y^{a})^{2}},~~~
G^{a\bar{b}}=-\delta^{ab}(z^{a}-\bar{z}^{a})^2=\delta^{ab}(2y^{a})^{2},~~~
\Gamma^{a}_{aa}=-\frac{2}{z^{a}-\bar{z}^{a}}.
\end{equation}
The superpotential can be written as
\begin{equation}
W=q_{0}+q_{a}z^{a}-p^{1}z^{2}z^{3}-p^{2}z^{1}z^{3}-p^{3}z^{1}z^{2}+p^{0}z^{1}z^{2}z^{3}.
\end{equation}

Consider a static, spherically symmetric four-dimensional spacetime
with the metric
\begin{equation}
ds^{2}=-e^{2U(\tau)}dt^{2}+e^{-2U(\tau)}d\vec{x}^{2},
\end{equation}
where $\tau=1/|\vec{x}|$. Note that the horizon locates at $\tau=0$
and the asymptotic infinity tends to $\tau\rightarrow\infty$. The
effective Lagrangian describing the system is
\begin{equation}
\mathcal{L}_{eff}=\dot{U}^{2}+G_{a\bar{b}}\dot{z}^{a}\dot{\bar{z}}^{\bar{b}}+e^{2U}V_{\rm
BH},
\end{equation}
where the dot denotes differentiation with respect to $\tau$. The
term $V_{\rm BH}$ in the above effective Lagrangian is the so-called
``effective potential'', which is given by
\begin{equation}
V_{BH}=|DZ|^{2}+|Z|^{2}=G^{a\bar{b}}(D_{a}Z)(\bar{D}_{\bar{b}}\bar{Z})+Z\bar{Z}.
\end{equation}
$Z$ is the central charge of the SUSY algebra and is expressed as
$Z=e^{K/2}W$ in our case. The covariant derivative of the central
charge is
\begin{equation}
D_{a}Z=e^{K/2}[\partial_{a}+(\partial_{a}K)]W.
\end{equation}

One can obtain the equations of motion by varying the above
effective Lagrangian.
\begin{equation}
\label{2eq15} \ddot{U}=e^{2U}V_{BH},
\end{equation}
\begin{equation}
\label{2eq16}
\ddot{z}^{a}+\Gamma^{a}_{bc}\dot{z}^{b}\dot{z}^{c}=e^{2U}G^{a\bar{b}}\partial_{\bar{b}}V_{\rm
BH}.
\end{equation}
In addition, there is a Hamiltonian constraint on the solutions
\begin{equation}
\label{2eq17}
\dot{U}^{2}+G_{a\bar{b}}\dot{z}^{a}\dot{\bar{z}}^{\bar{b}}-e^{2U}V_{BH}=c^{2},
\end{equation}
where $c^{2}=0$ for extremal black holes.

When the black hole in the solution is extremal, the values of the
moduli $z^{a}$ will be fixed at the horizon, irrespective of their
values at asymptotic infinity. The attractor values can be obtained
by minimizing the effective potential, either directly or via the
attractor equations. The entropy of the extremal black hole, whether
BPS or not, is given by the effective potential evaluated at the
extremum:
\begin{equation}
S=\frac{A}{4}=\pi V_{\rm BH}\mid_{\rm ext}.
\end{equation}

\subsection{The D0-D4 solution with complex moduli}
The simplest solution of the D0-D4 system is the case of a
D0-D4-D4-D4 black hole without B-fields, where $q_{0}>0$ and
$p^{a}>0$ but $p^{0}=q_{a}=0$, which results in a BPS configuration.
The solution to the effective Lagrangian is
\begin{equation}
e^{-4U}=4H_{0}H^{1}H^{2}H^{3},
\end{equation}
\begin{equation}
z^{a}=-i\sqrt{\frac{2H_{0}H^{a}}{s_{abc}H^{b}H^{c}}},
\end{equation}
where $s_{abc}=|\epsilon_{abc}|$ and the harmonic functions are
given as follows
\begin{equation}
H^{a}=\frac{1}{\sqrt{2}}+p^{a}\tau,~~~H_{0}=\frac{1}{\sqrt{2}}+q_{0}\tau.
\end{equation}
The attractor values of the moduli become
\begin{equation}
z^{a}=-i\sqrt{\frac{2q_{0}p^{a}}{s_{abc}p^{b}p^{c}}}.
\end{equation}

One can obtain a non-BPS solution by simply analytic continuation
from the BPS case, that is, we assume $q_{0}<0, p^{a}>0$. Thus the
harmonic functions turn out to be
\begin{equation}
H^{a}=\frac{1}{\sqrt{2}}+p^{a}\tau,~~~H_{0}=-\frac{1}{\sqrt{2}}+q_{0}\tau,
\end{equation}
and we can write down the solution
\begin{equation}
e^{-4U}=|4H_{0}H^{1}H^{2}H^{3}|,
\end{equation}
\begin{equation}
z^{a}=-i\sqrt{\frac{-2H_{0}H^{a}}{s_{abc}H^{b}H^{c}}}.
\end{equation}
The attractor values of the moduli
\begin{equation}
z^{a}=-i\sqrt{\frac{-2q_{0}p^{a}}{s_{abc}p^{b}p^{c}}}.
\end{equation}

The authors of~\cite{GLS} generalized the simple non-BPS solution to
situations where the asymptotic moduli are more general and/or there
are more charges present. In particular, they normalized the
asymptotic volume moduli so that $y^{a}|_{\infty}=1$ but kept the
asymptotic B-fields $x^{a}_{\infty}=B^{a}$ as free variables. They
obtained the following solution by solving the equations of motion
directly,
\begin{equation}
e^{-4U}=-4H_{0}H^{1}H^{2}H^{3}-B^{2},
\end{equation}
\begin{equation}
z^{a}=\frac{B-ie^{-2U}}{s_{abc}H^{b}H^{c}},
\end{equation}
where the harmonic functions are given by
\begin{equation}
H^{a}=\frac{1}{\sqrt{2}}+p^{a}\tau,~~~H_{0}=-\frac{1}{\sqrt{2}}(1+B^{2})+q_{0}\tau.
\end{equation}


\section{Exact Solution of the D2-D6 System}\label{sec:num}

The exact solutions to the D0-D4 system with general complex moduli
were obtained in~\cite{HK} and~\cite{GLS} by solving the equations
of motion directly. However, it would be more difficult to carry out
similar procedures for the D2-D6 system due to the cubic term
involving the D6 charge $p^{0}$ in the superpotential. However, the
exact solutions to the D2-D6 system with purely imaginary moduli
were obtained in~\cite{kal2}. The basic idea was that one could take
the horizon values of the moduli and replace the charges with
corresponding harmonic functions. Then one could check if the ansatz
satisfies the equations of motion as well as the constraint.
Fortunately after a somewhat more involved calculation one found
that the ansatz satisfied all the requirements. The result was
manifestly dual to the known D0-D4 system.

Inspired by such a method, we make a similar ansatz for the moduli
of the D2-D6 system and find that the ansatz also satisfies the
equations of motion and the constraint. Furthermore, the solution to
the D2-D6 system is also manifestly dual to the solution obtained
in~\cite{GLS}.

\subsection{The solutions to the D2-D6 system with purely imaginary moduli}\label{sec:exact_sol}
In this subsection we will list the main results of~\cite{kal2},
which leads to our ansatz. The superpotential for the D2-D6 system
is
\begin{equation}
\label{3eq1}
W=q_{a}z^{a}+p^{0}z^{1}z^{2}z^{3}.
\end{equation}
If $p^{0}q_{1}q_{2}q_{3}<0$, the resulting configuration is BPS,
while $p^{0}q_{1}q_{2}q_{3}>0$ corresponds to non-BPS configuration.
The attractor values for the moduli in the non-BPS case are given by
\begin{equation}
z^{1}=-i\sqrt{\frac{q_{2}q_{3}}{p^{0}q_{1}}},~~~z^{2}=
-i\sqrt{\frac{q_{1}q_{3}}{p^{0}q_{2}}},~~~z^{3}=-i\sqrt{\frac{q_{1}q_{2}}{p^{0}q_{3}}}.
\end{equation}

As pointed out in~\cite{kal2}, for both BPS and non-BPS STU black
holes, we can find solutions to the equations of motion for the
moduli, allowing us to obtain their values everywhere. Such
solutions can be obtained by replacing the charges in the attractor
values of the moduli with the corresponding harmonic functions.
Furthermore, we have to check that if such solutions satisfy the
equations of motion and the constraint.

For the D2-D6 case, after replacing the charges with harmonic
functions, the solutions turn out to be
\begin{equation}
e^{-2U}=2\sqrt{H^{0}H_{1}H_{2}H_{3}},
\end{equation}
\begin{equation}
z^{1}=-i\sqrt{\frac{H_{2}H_{3}}{H^{0}H_{1}}},~~~
z^{2}=-i\sqrt{\frac{H_{1}H_{3}}{H^{0}H_{2}}},~~~
z^{3}=-i\sqrt{\frac{H_{1}H_{2}}{H^{0}H_{3}}}.
\end{equation}
Note that the above equations can also be expressed as
\begin{equation}
z^{1}=-i\frac{e^{-2U}}{2H^{0}H_{1}},~~~
z^{2}=-i\frac{e^{-2U}}{2H^{0}H_{2}},~~~
z^{3}=-i\frac{e^{-2U}}{2H^{0}H_{3}}.
\end{equation}
 They proved that such an ansatz did satisfy the
equations of motion and the constraint by working out the terms in
these equations explicitly.

\subsection{The solutions to the D2-D6 system with general complex moduli}
In this subsection we will show that our ansatz satisfies the
equations of motion and the constraint of the D2-D6 system.
Furthermore, the exact solution is also manifestly dual to the
solution obtained in~\cite{GLS}. We would like to set
$z^{a}=x^{a}-iy^{a}$ for convenience.

The effective Lagrangian becomes
\begin{eqnarray}
\mathcal{L}_{eff}&=&(\dot{U})^{2}+G_{a\bar{b}}\dot{z}^{a}\dot{\bar{z}}^{\bar{b}}+e^{2U}V_{\rm
BH}\nonumber\\
&=&(\dot{U})^{2}+\frac{1}{4}\sum\limits^{3}_{a=1}\frac{(\dot{x}^{a})^{2}+(\dot{y}^{a})^{2}}{(y^{a})^{2}}
+e^{2U}V_{\rm BH}.
\end{eqnarray}
The expression for the effective potential is given by
\begin{eqnarray}
V_{\rm BH}&=&|DZ|^{2}+|Z|^{2}\nonumber\\
&=&G^{a\bar{b}}D_{a}Z\bar{D}_{\bar{b}}\bar{Z}+Z\bar{Z}\nonumber\\
&=&G^{a\bar{b}}(e^{K/2}(\partial_{a}+\partial_{a}K)W)(\overline{e^{K/2}(\partial_{b}+\partial_{b}K)W})+e^{K}W\overline{W}.
\end{eqnarray}
We can obtain the following explicit expression for the effective
potential by making use of~(\ref{2eq8}),~(\ref{2eq9})
and~(\ref{3eq1}),
\begin{eqnarray}
\label{3eq8} V_{\rm
BH}&=&\frac{1}{2y^{1}y^{2}y^{3}}\{q_{1}^{2}[(x^{1})^{2}+(y^{1})^{2}]+q_{2}^{2}[(x^{2})^{2}+(y^{2})^{2}]
+q_{3}^{2}[(x^{3})^{2}+(y^{3})^{2}]\nonumber\\&
&+(p^{0})^{2}[(x^{1})^{2}+(y^{1})^{2}][(x^{2})^{2}+(y^{2})^{2}]
[(x^{3})^{2}+(y^{3})^{2}]
+2p^{0}q_{1}x^{2}x^{3}[(x^{1})^{2}+(y^{1})^{2}]\nonumber\\&
&+2p^{0}q_{2}x^{1}x^{3}[(x^{2})^{2}+(y^{2})^{2}]
+2p^{0}q_{3}x^{1}x^{2}[(x^{3})^{2}+(y^{3})^{2}]\nonumber\\
&
&+2q_{1}q_{2}x^{1}x^{2}+2q_{1}q_{3}x^{1}x^{3}+2q_{2}q_{3}x^{2}x^{3}\}.
\end{eqnarray}
Then the equations of motion and the constraint turn out to be
\begin{equation}
\label{3eq9}
\ddot{U}=e^{2U}V_{\rm BH},
\end{equation}
\begin{eqnarray}
\label{3eq10}
\frac{1}{2}\frac{d}{d\tau}[\frac{\dot{x}^{1}}{(y^{1})^{2}}]&=&e^{2U}\frac{\partial
V_{\rm BH}}{\partial x^{1}},\nonumber\\
\frac{1}{2}\frac{d}{d\tau}[\frac{\dot{x}^{2}}{(y^{2})^{2}}]&=&e^{2U}\frac{\partial
V_{\rm BH}}{\partial x^{2}},\nonumber\\
\frac{1}{2}\frac{d}{d\tau}[\frac{\dot{x}^{3}}{(y^{3})^{2}}]&=&e^{2U}\frac{\partial
V_{\rm BH}}{\partial x^{3}},
\end{eqnarray}
\begin{eqnarray}
\label{3eq11}
\frac{1}{2}\frac{d}{d\tau}[\frac{\dot{y}^{1}}{(y^{1})^{2}}]+
\frac{1}{2(y^{1})^{3}}[(\dot{x}^{1})^{2}+(\dot{y}^{1})^{2}]&=&e^{2U}\frac{\partial
V_{\rm BH}}{\partial y^{1}},\nonumber\\
\frac{1}{2}\frac{d}{d\tau}[\frac{\dot{y}^{2}}{(y^{2})^{2}}]+
\frac{1}{2(y^{2})^{3}}[(\dot{x}^{2})^{2}+(\dot{y}^{2})^{2}]&=&e^{2U}\frac{\partial
V_{\rm BH}}{\partial y^{2}},\nonumber\\
\frac{1}{2}\frac{d}{d\tau}[\frac{\dot{y}^{3}}{(y^{3})^{2}}]+
\frac{1}{2(y^{3})^{3}}[(\dot{x}^{3})^{2}+(\dot{y}^{3})^{2}]&=&e^{2U}\frac{\partial
V_{\rm BH}}{\partial y^{3}}.
\end{eqnarray}

Assume the solutions take the following form
\begin{equation}
\label{3eq12} e^{-4U}=4H^{0}H_{1}H_{2}H_{3}-c^{2},
\end{equation}
\begin{equation}
z^{1}=x^{1}-i\frac{e^{2U}}{2H^{0}H_{1}},~~~z^{2}=x^{2}-i\frac{e^{2U}}{2H^{0}H_{2}},
~~~z^{3}=x^{3}-i\frac{e^{2U}}{2H^{0}H_{3}},
\end{equation}
\begin{equation}
H^{0}=a^{0}+p^{0}\tau,~~~H_{1}=a_{1}+q_{1}\tau,~~~H_{2}=a_{2}+q_{2}\tau,~~~H_{3}=a_{3}+q_{3}\tau,
\end{equation}
where $c$, $a^{0}$ and $a_{i}$(i=1,2,3) are numerical constants. Let
us solve~(\ref{3eq9}) first. From~(\ref{3eq12}) we have the
following expression
\begin{eqnarray}
\label{3eq15}
\ddot{U}&=&4e^{8U}(p^{0}H_{1}H_{2}H_{3}+q_{1}H^{0}H_{2}H_{3}+q_{2}H^{0}H_{1}H_{3}
+q_{3}H^{0}H_{1}H_{2})^{2}\nonumber\\
&
&-e^{4U}(2p^{0}q_{1}H_{2}H_{3}+2p^{0}q_{2}H_{1}H_{3}+2p^{0}q_{3}H_{1}H_{2}\nonumber\\
& &
~~~+2q_{1}q_{2}H^{0}H_{3}+2q_{1}q_{3}H^{0}H_{2}+2q_{2}q_{3}H^{0}H_{1}).
\end{eqnarray}
Thus we can expand both $\ddot{U}$ and $e^{2U}V_{\rm BH}$
using~(\ref{3eq8}) and~(\ref{3eq15}) then compare the corresponding
terms to find the solutions to $x^{1}, x^{2}, x^{3}$. Consider the
$q^{2}_{1}$ term for example, we obtain
\begin{equation}
4e^{8U}(H^{0}H_{2}H_{3})^{2}=\frac{e^{2U}}{2y^{1}y^{2}y^{3}}[(x^{1})^{2}+(y^{1})^{2}]
\end{equation}
Solving this equation gives
\begin{equation}
x^{1}=\frac{c}{2H^{0}H_{1}}.
\end{equation}
The solutions to $x^{2}$ and $x^{3}$ can be obtained in a similar
way, which gives
\begin{equation}
x^{2}=\frac{c}{2H^{0}H_{2}},~~~x^{3}=\frac{c}{2H^{0}H_{3}}.
\end{equation}
One can check that the above solutions solve~(\ref{3eq9})
completely.

The next task is to check if the above solutions satisfy the
remaining equations of motion and the constraint. Due to the cyclic
symmetries of $x^{i}$ and $y^{i}$ appear in $V_{\rm BH}$, it is
necessary to check the first two equations in~(\ref{3eq10})
and~(\ref{3eq11}). The left hand side of the first equation
in~(\ref{3eq10}) gives
\begin{equation}
\frac{1}{2}\frac{d}{d\tau}[\frac{\dot{x}^{1}}{(y^{1})^{2}}]=
-2cp^{0}q_{1}e^{4U}-4c(p^{0}H_{1}+q_{1}H^{0})e^{4U}\dot{U}.
\end{equation}
Note that
\begin{equation}
\label{3eq20}
\dot{U}=-e^{4U}(p^{0}H_{1}H_{2}H_{3}+q_{1}H^{0}H_{2}H_{3}+q_{2}H^{0}H_{1}H_{3}+q_{3}H^{0}H_{1}H_{2})
\end{equation}
and
\begin{eqnarray}
\frac{\partial V_{\rm BH}}{\partial
x^{1}}&=&\frac{1}{2y^{1}y^{2}y^{3}}\{2q^{2}_{1}+2q_{1}q_{2}x_{2}+2q_{1}q_{3}x_{3}+2(p^{0})^{2}x^{1}[(x^{2})^{2}
+(y^{2})^{2}][(x^{3})^{2} +(y^{3})^{2}]\nonumber\\
& &+4p^{0}q_{1}x^{1}x^{2}x^{3}+2p^{0}q_{2}[(x^{2})^{2}
+(y^{2})^{2}]x^{3}+2p^{0}q_{3}[(x^{3})^{2} +(y^{3})^{2}]x^{2}\}.
\end{eqnarray}
Substituting the expressions of $x^{1}$ and $y^{1}$ to the above
equations, one can find that both sides match after a lengthy
calculation.

Subsequently we rewrite the first equation of~(\ref{3eq11}) as
follows:
\begin{equation}
\label{3eq22}
\frac{1}{2}\frac{\ddot{y}^{1}}{(y^{1})^{2}}+\frac{1}{2(y^{1})^{3}}(\dot{x}^{1})^{2}-
\frac{1}{2(y^{1})^{3}}(\dot{y}^{1})^{2}=e^{2U}\frac{\partial V_{\rm
BH}}{\partial y^{1}}.
\end{equation}
Note that
\begin{equation}
\label{3eq23}
\dot{x}^{1}=-x^{1}(\frac{p^{0}}{H^{0}}+\frac{q_{1}}{H_{1}}),
\end{equation}
\begin{equation}
\label{3eq24}
\dot{y}^{1}=-y^{1}(2\dot{U}+\frac{p^{0}}{H^{0}}+\frac{q_{1}}{H_{1}}),
\end{equation}
\begin{equation}
\ddot{y}^{1}=y^{1}[4\dot{U}^{2}+4\dot{U}(\frac{p^{0}}{H^{0}}+\frac{q_{1}}{H_{1}})+
(\frac{p^{0}}{H^{0}}+\frac{q_{1}}{H_{1}})^{2}-2\ddot{U}+\frac{(p^{0})^{2}}{(H^{0})^{2}}+
\frac{q_{1}^{2}}{H_{1}^{2}}].
\end{equation}
Thus the left hand side of~(\ref{3eq22}) can be given as
\begin{equation}
\frac{1}{2}\frac{\ddot{y}^{1}}{(y^{1})^{2}}+\frac{1}{2(y^{1})^{3}}(\dot{x}^{1})^{2}-
\frac{1}{2(y^{1})^{3}}(\dot{y}^{1})^{2}=-\frac{\ddot{U}}{y^{1}}+\frac{1}{2y^{1}}[\frac{(p^{0})^{2}}{(H^{0})^{2}}+
\frac{q_{1}^{2}}{H_{1}^{2}}]+\frac{(x^{1})^{2}}{2(y^{1})^{3}}(\frac{p^{0}}{H^{0}}+\frac{q_{1}}{H_{1}})^{2}
\end{equation}
while
\begin{equation}
\frac{\partial V_{\rm BH}}{\partial y^{1}}=-\frac{V_{\rm
BH}}{y^{1}}+\frac{1}{2y^{1}y^{2}y^{3}}\{2q^{2}_{1}y^{1}+2(p^{0})^{2}y^{1}[(x^{2})^{2}
+(y^{2})^{2}][(x^{3})^{2}
+(y^{3})^{2}]+4p^{0}q_{1}x^{2}x^{3}y^{1}\}.
\end{equation}
One can find that our solutions also satisfy this equation by making
use of the equation of motion~(\ref{3eq9}).

Finally, we have to check the constraint~(\ref{2eq17}), which can be
rewritten as follows
\begin{equation}
\label{3eq28}
(\dot{U})^{2}+\frac{1}{4}\sum\limits^{3}_{a=1}\frac{(\dot{x}^{a})^{2}+(\dot{y}^{a})^{2}}{(y^{a})^{2}}
=e^{2U}V_{\rm BH}.
\end{equation}
By making use of~(\ref{3eq20}),~(\ref{3eq23}),~(\ref{3eq24}) as well
as the cyclic permutations of the last two equations, one can find
that the left hand side of~(\ref{3eq28}) can be simplified
dramatically,
\begin{equation}
(\dot{U})^{2}+\frac{1}{4}\sum\limits^{3}_{a=1}\frac{(\dot{x}^{a})^{2}+(\dot{y}^{a})^{2}}{(y^{a})^{2}}=\ddot{U}.
\end{equation}
Thus the constraint is also satisfied according to the equation of
motion~(\ref{3eq9}).

Now we would like to summarize our main result. The solution to the
non-BPS D2-D6 system can be expressed as
\begin{equation}
e^{-4U}=4H^{0}H_{1}H_{2}H_{3}-c^{2},
\end{equation}
\begin{equation}
z^{1}=\frac{c}{2H^{0}H_{1}}-i\frac{e^{2U}}{2H^{0}H_{1}},~~~z^{2}=\frac{c}{2H^{0}H_{2}}-i\frac{e^{2U}}{2H^{0}H_{2}},
~~~z^{3}=\frac{c}{2H^{0}H_{3}}-i\frac{e^{2U}}{2H^{0}H_{3}},
\end{equation}
\begin{equation}
H^{0}=a^{0}+p^{0}\tau,~~~H_{1}=a_{1}+q_{1}\tau,~~~H_{2}=a_{2}+q_{2}\tau,~~~H_{3}=a_{3}+q_{3}\tau.
\end{equation}
It can be easily seen that the above solutions have a manifestly
dual form with respect to the solution obtained in~\cite{GLS}. The
numerical constants are left undetermined due to subtle points which
will be discussed in the last section. One can see that the moduli
exhibit the same attractor values at the horizon as those of the
simple D2-D6 non-BPS black hole. The entropy is given by
\begin{equation}
S=\pi V|_{ext}=2\pi\sqrt{p^{0}q_{1}q_{2}q_{3}},
\end{equation}
which is also the same as that of the simple D2-D6 case.
\section{Adding D2 Charges to D0-D4 System}
In this section we obtain new solutions carrying D0-D2-D4 charges by
transforming the original D0-D4 solution in~\cite{GLS} under
$SL(2,{\bf Z})^{3}$ duality.
\subsection{U-duality for STU black holes}
The symplectic structure of $\mathcal{N}=2$ supergravity admits a
symplectic invariant $I_{1}$, which is given by
\begin{equation}
I_{1}=|Z|^{2}+|DZ|^{2}
\end{equation}
$I_{1}$ becomes a function of charges when restricted to STU model,
which is given by $I_{1}=\sqrt{|\mathcal{W}(\Gamma)}|$, where
\begin{eqnarray}
\mathcal{W}(\Gamma)&=&4((p^{1}q_{1})(p^{2}q_{2})+(p^{1}q_{1})(p^{3}q_{3})+(p^{2}q_{2})(p^{3}q_{3}))\nonumber\\
&
&-(p^{\Lambda}q_{\Lambda})^{2}-4p^{0}q_{1}q_{2}q_{3}+4q_{0}p^{1}p^{2}p^{3}
\end{eqnarray}
and $\Gamma=(p^{\Lambda}~q_{\Lambda}), \Lambda=0,1,2,3$. The
symplectic invariance of special geometry ensures that the
Lagrangian has an $Sp(8, \bf Z)$ symmetry, which reduces to
$[SL(2,{\bf Z})]^{3}$ at the level of the the equations of motion.
Given an $SL(2,\bf Z)$ matrix
$$
\left(
\begin{array}{cc}
a&b\\
c&d
\end{array}
\right),
$$with $ad-bc=1$, the moduli change as
\begin{equation}
\label{4eq3} \tilde{z}^{a}=\frac{az^{a}+b}{cz^{a}+d}.
\end{equation}

In~\cite{kal2}, the authors changed the charges appeared in the
attractor values of the moduli to the corresponding harmonic
functions. Consider the spherically symmetric, static metric ansatz:
\begin{equation}
ds^{2}=-e^{2U}dt^{2}+e^{-2U}d\vec{x}^{2},
\end{equation}
where the metric is given by
\begin{equation}
e^{-2U}=\sqrt{|\mathcal{W({\bf H}})|}.
\end{equation}
In order to obtain general (D0, D2, D4, D6) system from their D2-D6
solution, they took a specific element of $[SL(2,{\bf Z})]^{3}$ and
obtained the general solutions via duality transformations. We will
find the D0-D2-D4 solution in a similar way in the next section.
\subsection{The D0-D2-D4 solution}
First consider the general superpotential~(\ref{2eq6}). For a
configuration carrying D0-D4 charges
\begin{equation}
W=q_{0}-3D_{AB}z^{A}z^{B},
\end{equation}
where $D_{AB}\equiv D_{ABC}p^{C}$. If we add D2 charges to the
system, the superpotential turns out to be
\begin{equation}
W=q_{0}+q_{A}z^{A}-3D_{AB}z^{A}z^{B},
\end{equation}
However, if we do the following transformations
\begin{equation}
\label{4eq7}
\hat{q}_{0}=q_{0}+\frac{1}{12}D^{AB}q_{A}q_{B},~~~\hat{z}^{A}=z^{A}-\frac{1}{6}D^{AB}q_{B},
\end{equation}
where $D^{AB}\equiv (D_{AB})^{-1}$. Then the D0-D2-D4 superpotential
becomes
\begin{equation}
W=\hat{q}_{0}-3D_{AB}\hat{z}^{A}\hat{z}^{B},
\end{equation}
which has the same form as the D0-D4 case.

Now let us specialize to the STU model. The matrices $D_{ab}$ and
$D^{ab}$ are given explicitly as follows:
\begin{equation}
D_{ab}=\frac{1}{6}\left(
\begin{array}{ccc}
0&p^{3}&p^{2}\\
p^{3}&0&p^{1}\\
p^{2}&p^{1}&0
\end{array}
\right)
\end{equation}
The inverse matrix
\begin{equation}
D^{ab}=\left(
\begin{array}{ccc}
-\frac{3p^{1}}{p^{2}p^{3}}&\frac{3}{p^{3}}&\frac{3}{p^{2}}\\
\frac{3}{p^{3}}&-\frac{3p^{2}}{p^{1}p^{3}}&\frac{3}{p^{1}}\\
\frac{3}{p^{2}}&\frac{3}{p^{1}}&-\frac{3p^{3}}{p^{1}p^{2}}
\end{array}
\right)
\end{equation}
Then from~(\ref{4eq7}) we have the following transformations
\begin{equation}
\label{4eq11}
z^{\prime a}=z^{a}+k^{a},
\end{equation}
where the quantities with primes belong to the D0-D2-D4 system and
the quantities without primes belong to the original D0-D4 system
from now on. $k^{a}$ can be written as
\begin{eqnarray}
k^{1}&\equiv&\frac{1}{6}D^{1b}q_{b}=\frac{q_{2}}{2p^{3}}+\frac{q_{3}}{2p^{2}}-\frac{p^{1}q_{1}}{2p^{2}p^{3}},\nonumber\\
k^{2}&\equiv&\frac{1}{6}D^{2b}q_{b}=\frac{q_{1}}{2p^{3}}+\frac{q_{3}}{2p^{1}}-\frac{p^{2}q_{2}}{2p^{1}p^{3}},\nonumber\\
k^{3}&\equiv&\frac{1}{6}D^{3b}q_{b}=\frac{q_{1}}{2p^{2}}+\frac{q_{2}}{2p^{1}}-\frac{p^{3}q_{3}}{2p^{1}p^{2}}.
\end{eqnarray}
Note that the charge configuration of the D0-D4 system is expressed
as $(\hat{q}_{0},0,0,p^{a})$ while for the D0-D2-D4 system we have
$(q_{0},q_{a}, 0, p^{\prime a})$.

According to~(\ref{4eq3}) and~(\ref{4eq11}), we can write down the
$[SL(2,{\bf Z})]^{3}$ matrices
\begin{equation}
\label{4eq14} M_{1}=\left(
\begin{array}{cc}
1&k^{1}\\
0&1
\end{array}
\right),~~~ M_{2}=\left(
\begin{array}{cc}
1&k^{2}\\
0&1
\end{array}
\right),~~~ M_{3}=\left(
\begin{array}{cc}
1&k^{3}\\
0&1
\end{array}
\right).~~~
\end{equation}
Our task is to generalize new solutions by making use of these
$[SL(2,{\bf Z})]^{3}$ matrices.

Take the same notations as those in~\cite{GLS}
\begin{eqnarray}
p^{0}&=&a_{111},~~q_{0}=-a_{000},~~p^{1}=a_{011},~~q_{1}=a_{100},\nonumber\\
p^{2}&=&a_{101},~~q_{2}=a_{010},~~~p^{3}=a_{110},~~~q_{3}=a_{001},
\end{eqnarray}
which transform as
\begin{equation}
a^{\prime}_{i^{\prime}j^{\prime}k^{\prime}}={(M_{1})_{i^{\prime}}}^{i}{(M_{2})_{j^{\prime}}}^{j}
{(M_{3})_{k^{\prime}}}^{k}a_{ijk}~~~i,j,k=0,1.
\end{equation}
Using~(\ref{4eq14}), one can check that
\begin{equation}
(\hat{q_{0}},0,0,p^{a})\Rightarrow(q_{0},q_{a},0,p^{a}).
\end{equation}
Similarly, the harmonic functions transform as
\begin{equation}
\label{4eq18}
{H}^{\prime1}=H^{1},~~~H^{\prime2}=H^{2},~~~H^{\prime3}=H^{3}.
\end{equation}
\begin{eqnarray}
\label{4eq19}
H^{\prime}_{1}&=&k^{3}H^{2}+k^{2}H^{3}=\frac{1}{\sqrt{2}}(k^{2}+k^{3})+q_{1}\tau,\nonumber\\
H^{\prime}_{2}&=&k^{3}H^{1}+k^{1}H^{3}=\frac{1}{\sqrt{2}}(k^{1}+k^{3})+q_{2}\tau,\nonumber\\
H^{\prime}_{3}&=&k^{1}H^{2}+k^{2}H^{1}=\frac{1}{\sqrt{2}}(k^{1}+k^{2})+q_{3}\tau,\nonumber\\
\end{eqnarray}
\begin{eqnarray}
\label{4eq20}
H^{\prime}_{0}&=&H_{0}-k^{1}k^{2}H^{3}+k^{1}k^{3}H^{2}+k^{2}k^{3}H^{1}\nonumber\\
&=&\frac{1}{\sqrt{2}}[(1+B^{2})-(k^{1}k^{2}+k^{2}k^{3}+k^{1}k^{3})]+q_{0}\tau.
\end{eqnarray}
We can see that the duality invariant does not change indeed
\begin{eqnarray}
I_{1}(\Gamma)&=&4\hat{q}_{0}p^{1}p^{2}p^{3}\nonumber\\
&=&4q_{0}p^{1}p^{2}p^{3}-(p^{1})^{2}q^{2}_{1}-(p^{2})^{2}q^{2}_{2}-(p^{3})^{2}q_{3}^{2}\nonumber\\
& &+2q_{1}q_{2}
p^{1}p^{2}+2q_{1}q_{3}p^{1}p^{3}+2q_{2}q_{3}p^{2}p^{3}.
\end{eqnarray}
\begin{eqnarray}
I^{\prime}_{1}(\Gamma)&=&4q_{0}p^{1}p^{2}p^{3}-(p^{1}q_{1}+p^{2}q_{2}+p^{3}q_{3})^{2}+4(p^{1}q_{1}p^{2}q_{2}
+p^{1}q_{1}p^{3}q_{3}+p^{2}q_{2}p^{3}q_{3})\nonumber\\
&=&4q_{0}p^{1}p^{2}p^{3}-(p^{1})^{2}q^{2}_{1}-(p^{2})^{2}q^{2}_{2}-(p^{3})^{2}q_{3}^{2}\nonumber\\
& &+2q_{1}q_{2}
p^{1}p^{2}+2q_{1}q_{3}p^{1}p^{3}+2q_{2}q_{3}p^{2}p^{3}\nonumber\\
&=&I_{1}
\end{eqnarray}
Furthermore, we can see that $\mathcal{W}({\bf H})$ is also
invariant.
\begin{equation}
\mathcal{W}({\bf H})=4H_{0}H^{1}H^{2}H^{3}.
\end{equation}
\begin{eqnarray}
\mathcal{W}^{\prime}({\bf
H^{\prime}})&=&4((H^{\prime1}H^{\prime}_{1})(H^{\prime2}H^{\prime}_{2})+
(H^{\prime1}H^{\prime}_{1})(H^{\prime3}H^{\prime}_{3})+(H^{\prime2}H^{\prime}_{2})
(H^{\prime3}H^{\prime}_{3}))\nonumber\\
&
&-(H^{\prime1}H^{\prime}_{1}+H^{\prime2}H^{\prime}_{2}+H^{\prime3}H^{\prime}_{3})^{2}
+4H^{\prime}_{0}H^{\prime1}H^{\prime2}H^{\prime3}.
\end{eqnarray}
After substituting the expressions for $H^{\prime}$ harmonic
functions~(\ref{4eq18})-~(\ref{4eq20}), one can find that
\begin{equation}
\mathcal{W}^{\prime}({\bf H^{\prime}})=\mathcal{W}({\bf H}).
\end{equation}
Here the metric is given by
\begin{equation}
e^{-4U}=|\mathcal{W}({\bf H})|-c^{2}.
\end{equation} Thus
\begin{equation}
\label{4eq27} e^{-2U^{\prime}}=e^{-2U}.
\end{equation}
Now we have to check whether the new solution satisfies the
equations of motion. From~(\ref{4eq11}) and (\ref{4eq27}) we can see
that the left hand side of the equations of motion and the
constraint remain unchanged. Thus we just need to check if the
effective potential $V_{\rm BH}$ on the right hand side remains
invariant. It can be easily seen that
\begin{equation}
K^{\prime}=K,~~~W=W^{\prime},~~~D_{a}Z=D_{a\prime}Z^{\prime},
\end{equation}
after taking all the relevant formulae into account. Thus by the
definition of $V_{\rm BH}$, we have
\begin{equation}
V^{\prime}_{\rm BH}=V_{\rm BH}.
\end{equation}
Then our new solution also satisfies the equations of motion and the
constraint.

One can easily obtain the attractor values of the moduli
\begin{equation}
z^{\prime a}=k^{a}-i\sqrt{\frac{-2q_{0}p^{a}}{s_{abc}p^{b}p^{c}}},
\end{equation}
which is the same as those discussed in previous examples. The
entropy is given by
\begin{equation}
S_{\rm BH}=\pi V_{\rm
BH}|_{ext}=2\pi\sqrt{\hat{q}_{0}p^{1}p^{2}p^{3}},
\end{equation}
which also agrees with the previously known D0-D2-D4 entropy.
\section{Summary and Discussion}
In this note we obtain the non-BPS, extremal, spherically symmetric
black hole solutions of four-dimensional supergravity, carrying
D2-D6 and D0-D2-D4 charges. The D2-D6 solution contains general
complex moduli and is manifestly dual to the D0-D4 cousin appeared
in~\cite{GLS}. The D0-D2-D4 solution is obtained by $[SL(2,{\bf
Z)}]^{3}$ duality transformations from the D0-D4 solution. Both of
our solutions give the same attractor values of the moduli and the
same entropies as those of previously known examples carrying the
same charges. One may obtain new solutions carrying general (D0,
D2,D4,D6) charges from the known solutions by duality
transformations.

One subtle point is the determination of the numerical constants in
the D2-D6 solution. Of course one can take the same values as those
in~\cite{GLS}, that is,
\begin{equation}
c=B,~~~,a_{0}=\frac{1}{\sqrt{2}}(1+B^{2}),~~~a^{1}=a^{2}=a^{3}=\frac{1}{\sqrt{2}}.
\end{equation}
Thus one can obtain the mass by expanding the warp factor,
\begin{equation}
2G_{N}M_{\rm
non-BPS}=\frac{1}{\sqrt{2}}(|p^{0}|+\sum\limits_{a}q_{a}(1+B^{2})),
\end{equation}
which has a similar form as that given in~\cite{GLS}. This can be
interpreted as the sum of the masses of the D2 and D6-branes, which
also  exhibits a marginal bound state behavior. However, in such
cases the asymptotic values of the moduli are different,
\begin{equation}
z^{a}=\frac{B-i}{1+B^{2}},
\end{equation}
which means that the normalization of the asymptotic moduli should
be different from that in~\cite{GLS}.

Another interesting non-BPS configuration is the D0-D6 system, which
has been extensively studied in recent years, see
e.g.\cite{WT}~\cite{DM}\cite{FL}~\cite{EM}. The attractor mechanism
for D0-D6 Kaluza-Klein black holes has been discussed
in~\cite{Astefanesei:2006dd} using the entropy function formalism.
Since our solution is manifest dual to the D0-D4 solution, it will
be interesting to study the relations between our solution and the
D0-D6 system discussed in~\cite{GLS}.

The STU model has been discussed in~\cite{Levay:2007nm} in another
interesting way, that is, such a model can be tackled in the context
of quantum information theory. The use of this formalism expresses
the black hole potential in an especially elegant form as the norm
squared of a suitable tripartite entangled state. Then the
classification of solutions proceeds with analysing the charge codes
using some elements of quantum error correction. However, only
doubly extremal solutions were discussed in that paper for
illustration. So it would be interesting to extend similar analysis
using the more general solutions.

A further direction is to generalize the famous ``OSV''
conjecture~\cite{OSV}, which relates the partition function of BPS
black holes to the partition function of topological strings, to
non-BPS case. In a recent paper~\cite{SV}, Sarakin and Vafa pointed
out that there was some subtle points when generalizing the original
``OSV'' conjecture to non-BPS cases. Thus an extension of OSV that
can be applied simultaneously to both BPS and non-BPS black holes is
needed, which is more difficult to realize. However, the various
exact solutions of non-BPS black holes provide concrete examples for
testing their conjecture. We would like to study this problem in the
near future.

{\bf Note Added}: After the first version appearing on arXiv, we
were informed with~\cite{CCDOP}. In that interesting paper the
authors constructed interpolating solutions describing single-center
static extremal non-supersymetric black holes in four dimensional
$\mathcal{N}=2$ supergravity with cubic prepotentials. They derived
and solved the first-order flow equations for 5D rotating
electrically charged extremal black holes in a Taub-NUT geometry.
Then using the 4D/5D connections they obtained the corresponding 4D
solutions. One key point for these results was that the 5D geometry
was assumed to be a time fibration over a Hyper-K\"{a}hler base.
When the 4D prepotential contains a cubic term, the corresponding
solutions to the first-order flow equations are
\begin{equation}
e^{-4U}=\frac{4}{9}N(H_{A}f^{-1/2}X^{A})^{2}-c^{2},
\end{equation}
\begin{equation}
z^{A}=\frac{3}{2}(\frac{c+ie^{-2U}}{NH_{B}f^{-1/2}X^{B}})f^{-1/2}X^{A}.
\end{equation}
One can find that our D2-D6 solution agrees with their solution when
restricted to STU model\footnote{We thank G. Cardoso for pointing
this to us.}.

\bigskip \goodbreak \centerline{\bf Acknowledgements}
\noindent
We thank G. Cardoso and P. Levay for correspondence. The
work was supported in part by a grant from Chinese Academy of
Sciences, by NSFC under grants No. 10325525 and No. 90403029.



\end{document}